\documentclass[12pt,doublespace]{JHEP3}

\usepackage{epsfig}
\usepackage{amsmath}
\usepackage{xspace}
\usepackage{mathrsfs}

\numberwithin{equation}{section}

\def\varpi{t}

\def\({\left(}
\def\){\right)}
\def\[{\left[}
\def\]{\right]}
\def\hf{{1\over 2}}

\def\ba{\bar a}

\def\bZ{\bar Z}

\newcommand{\cD}{\ensuremath{\mathcal D}}

\newcommand{\cI}{\ensuremath{\mathcal I}}
\newcommand{\cZ}{\ensuremath{\mathcal Z}}
\newcommand{\cM}{\ensuremath{\mathcal M}}
\newcommand{\cN}{\ensuremath{\mathcal N}}
\newcommand{\cO}{\ensuremath{\mathcal O}}
\newcommand{\cH}{\ensuremath{\mathcal H}}

\newcommand{\cX}{\ensuremath{\mathcal X}}

\newcommand{\IR}{\ensuremath{\mathbb R}}
\newcommand{\IZ}{\ensuremath{\mathbb Z}}
\newcommand{\IN}{\ensuremath{\mathbb N}}

\newcommand{\half}{\ensuremath{\frac{1}{2}}}

\newcommand{\hk}{hyperk\"ahler\xspace}

\newcommand{\I}{{\mathrm i}}
\newcommand{\e}{{\mathrm e}}
\newcommand{\de}{\mathrm{d}}

\newcommand{\eps}{\epsilon}

\newcommand{\pa}{{\partial}}

\DeclareMathOperator{\Tr}{Tr}

\def\sgn{{\rm sgn\,}}

\def\bea{\begin{eqnarray}}
\def\eea{\end{eqnarray}}
\def\be{\begin{equation}}
\def\ee{\end{equation}}
\def\ba{\begin{align}}
\def\ea{\end{align}}
\def\bse{\begin{subequations}}
\def\ese{\end{subequations}}

\newcommand{\txi}{\tilde\xi}

\def\xp{x_+}
\def\xm{x_-}

\title{Wall-crossing made smooth}

\preprint{arXiv:15mm.nnnnn\\
CERN-PH-TH/2015-002
}

\author{Boris Pioline
\\

{\it CERN PH-TH,
Case C01600, CERN, CH-1211 Geneva 23, Switzerland}\\

{\it Sorbonne Universit\'es, UPMC Universit\'e Paris 6, UMR 7589, F-75005 Paris, France}\\

 {\it Laboratoire de Physique Th\'eorique et Hautes
Energies, CNRS UMR 7589, \\
Universit\'e Pierre et Marie Curie,
4 place Jussieu, F-75252 Paris cedex 05, France}

\vspace*{2mm} {\tt e-mail: \email{
 boris.pioline@cern.ch
}
} \vspace*{-3mm}

}

\abstract{In $D=4, \cN=2$ theories on $\IR^{3,1}$, the index receives contributions not only from single-particle BPS states, counted by the BPS indices, but also from multi-particle states made of BPS constituents. In a recent work  \cite{Alexandrov:2014wca}, a general formula 
expressing the index in terms of the BPS indices was proposed,  which is smooth across walls of marginal stability and reproduces the expected single-particle contributions. In this note, I analyze the two-particle contributions predicted by this formula, and show agreement with the 
spectral asymmetry of the continuum of scattering states in the supersymmetric 
quantum mechanics of two non-relativistic, mutually non-local dyons. 
This provides a physical justification for the error function profile used in the mathematics literature on indefinite theta series, and in the 
physics literature on black hole partition functions.
}

\begin{document}

\section{Introduction}

The recent work \cite{Alexandrov:2014wca} proposed a general formula for the index $\cI(R,u,C)$ in four-dimensional field theories on $\IR^{3,1}$ with  $\cN=2$ supersymmetry. This index can be understood as the partition function on $\IR^3$ times an Euclidean circle of radius $R$,
with periodic boundary conditions for fermions, chemical potentials $C$ conjugate to the electromagnetic
charge $\gamma$, and with an insertion of a suitable four-fermion vertex so as to saturate fermionic zero-modes. Equivalently, it can be defined as a trace
\be
\label{defWittenZ}
\cI(R,u,C) =
-\tfrac12\Tr_{\!\cH(u)} (-1)^{2J_3}(2J_3)^2  \sigma_\gamma\, e^{-2\pi R H-2\pi\I \langle \gamma, C \rangle},
\ee
over the full Hilbert space $\cH(u)$ of the theory on $\IR^3$ (here $J_3$ is the angular momentum operator around a fixed axis, and $\sigma_\gamma$ is a charge-dependent sign, satisfying the quadratic refinement property
$\sigma_\gamma \sigma_{\gamma'} = (-1)^{\langle \gamma,\gamma'\rangle} \sigma_{\gamma+\gamma'}$ ). 

Unlike the BPS indices $\Omega(\gamma,u)$, which count single-particle BPS states and exhibit discontinuities across walls of marginal stability, $\cI(R,u,C)$ is a smooth function of the Coulomb branch moduli $u$, away from the loci where additional states become massless. This is possible because $\cI(R,u,C)$ receives contributions not only from single-particle BPS states, but also from the continuum of multi-particle states. Indeed, while multi-particle states do not saturate the BPS bound $M=|Z_\gamma|$, the density 
of bosonic and fermionic states are not necessarily equal, as noted 
early on in \cite{Kaul:1983yt,Akhoury:1984pt} (see \cite{Troost:2010ud,Ashok:2011cy,
Harvey:2014nha} for recent discussions in the context of two-dimensional superconformal field theories).
Still, only multi-particle states made of BPS constituents can contribute, so one expects that the
index can be expressed in terms of the BPS indices $\Omega(\gamma,u)$.

In \cite{Alexandrov:2014wca}, using insight from the study of the \hk metric on the Coulomb branch in the theory
compactified down to three dimensions \cite{Gaiotto:2008cd}, and by analogy with a similar construction in 
the context of the hypermultiplet moduli space in string vacua \cite{Alexandrov:2008gh,Alexandrov:2009zh}, 
we proposed a general formula for 
the  index\footnote{Various refinements of this index have been introduced in 
\cite{Cecotti:2014wea}, but lie beyond the scope of this note.} 
\be
\label{defZc}
\cI(R,u,C)=- \frac{R^2}{2}\,\Im(\bar X^\Lambda F_\Lambda)+ 
 \frac{ R}{16\I\pi^2}
\sum_\gamma \Omega(\gamma,u)
\int_{\ell_\gamma} \frac{\text{d}t}{t}\left( t^{-1}  Z_\gamma  -t\bar Z_\gamma \right) \log\(1-\cX_\gamma(t)\).
\ee
where $\ell_{\gamma}$ are the BPS rays $\{ t \in \mathbb{C}^{\times}:\  Z_{\gamma}/t \in \I\IR^-\}$ , and $\cX_\gamma$ are the solutions to the system of integral equations \cite{Gaiotto:2008cd}
\be
{\cX_\gamma}={\cX_\gamma^{\text{sf}}}\,\exp\!\!\left[\sum_{\gamma'}
\frac{\Omega(\gamma',u)}{4\pi\I}\left<\gamma, \gamma'\right>
\!\! \int_{\ell_{\gamma'}} \!\!\!\! \frac{\text{d}t'}{t'}
\frac{t+t'}{t-t'}\log\(1- \cX_{\gamma'}(t')\)\right] ,
\label{TBA-HK}
\ee
with $\cX^{\text{sf}}_\gamma$ providing the `semi-flat', large $R$ approximation to 
$\cX_\gamma$,
\be
\label{defXsf}
\cX^{\text{sf}}_\gamma=\sigma_{\gamma}\,
e^{-\pi \I R \left(t^{-1} Z_\gamma - t \bZ_\gamma \right)-2\pi \I\left<\gamma, C \right>}.
\ee
The $\cX_\gamma$'s are holomorphic functions on the twistor space $\cZ$ of the Coulomb branch $\cM_3(R)$, which provide canonical Darboux coordinates for the holomorphic symplectic structure
on $\cZ$. They can also be understood as vevs of certain infrared line operators  \cite{Gaiotto:2010be}. In the limit $R\to\infty$, a formal solution to the system \eqref{TBA-HK} is obtained by
substituting $\cX_\gamma\to \cX^{\text{sf}}_\gamma$ on the r.h.s. and iterating. This leads
an expansion of the form
\be
\cX_\gamma = \cX_\gamma^{\text{sf}}\, \exp\left[ \sum_T 
\prod_{(i,j)\in T_1} \langle \alpha_i, \alpha_j\rangle\, \prod_{i\in T_0} \overline{\Omega}(\alpha_i,u)\,
g_T \right]
\ee
where $T$ runs over  trees decorated by charges $\alpha_i$ such that $\gamma=\sum \alpha_i$,
 $g_T$ are certain iterated contour integrals \cite{Gaiotto:2008cd,stoppa2014}, and 
 $\overline{\Omega}(\gamma,u)=\sum_{d|\gamma} \tfrac{1}{d^2}  \Omega(\gamma/d,u)$ are the `rational BPS indices' \cite{Joyce:2008pc,ks,Manschot:2010qz}, which arise from expanding
 the log in \eqref{TBA-HK}. Substituting in \eqref{defZc}, one obtains a formal expansion 
 \be
 \cI = \cI^{(0)} + {\sum\limits_\gamma} \, \cI_\gamma^{(1)} + {\sum\limits_{\gamma,\gamma'}} \, \cI_{\gamma,\gamma'}^{(2)} + \dots
 \ee
 where $\cI^{(0)}$ stands for the first term in \eqref{defZc}, while  $\cI_{\gamma_1,\dots,\gamma_n}^{(n)}$, proportional to $\overline{\Omega}(\gamma_1,u)\dots \overline{\Omega}(\gamma_n,u)$
is interpreted as the contribution of a multi-particle state of charge $\{\gamma_1,\dots \gamma_n\}$
to the index. In particular, the one-particle contribution is obtained
by replacing $\cX_\gamma\to\cX_{\gamma}^{\rm sf}$ in \eqref{defZc}, leading to
\be
\label{Z1inst}
\cI_\gamma^{(1)}=  \frac{R}{4\pi^2} 
\sigma_\gamma\,\overline{\Omega}(\gamma,u)\,
|Z_\gamma|\,
K_1(2\pi  R |Z_\gamma|)\, e^{-2\pi\I  \langle \gamma,C \rangle}
\ee
In \cite{Alexandrov:2014wca}, we matched this result with the index 
of a relativistic particle of charge $\gamma$ and
mass $|Z_\gamma|$. To define the index, we regulated the infrared divergences by switching on a chemical potential
$\theta$ for the rotations $J_3$ in the $xy$ plane, restricting the $z$ direction to a finite interval of length $L$,
and removing the regulators as follows,
\be
\label{Z1vsTr}
\cI_\gamma^{(1)}= 2R\!\lim\limits_{\substack{\theta\to 2\pi\\ L\to\infty}} \pa_{\theta}^2 \left[ \frac{\sin^2(\theta/2)}{\pi L}
\,
\Tr_{\cH_1(u)}\! \(\sigma_\gamma\,e^{-2\pi R H+\I\theta J_3-2\pi\I \langle \gamma,C\rangle}\)
\right]\! .
\ee
where $\cH_1(u)$ is the one-particle Hilbert space. The same regulator should then be used
to define the full index \eqref{defWittenZ}. Our aim in this note is to perform a similar check for the two-particle contribution.

\section{Two-particle contribution to the index}
According to the conjecture \eqref{defZc}, the contribution of a two-particle state with charges
$\{\gamma,\gamma'\}$ to the index is obtained
by inserting the one-particle approximation to \eqref{TBA-HK} in \eqref{defZc},
\be
\cI^{(2)}_{\gamma,\gamma'}= -\frac{ R}{64\pi^3} {\sum\limits_{\gamma,\gamma'}}\, 
\langle \gamma,\gamma'\rangle\, 
\bar\Omega(\gamma)\, \bar\Omega(\gamma') 
\, \int_{\ell_\gamma} \frac{\text{d}t}{t} \,
\int_{\ell_\gamma'} \frac{\text{d}t'}{t'}\frac{t+t'}{t-t'}
\left( t^{-1} Z_\gamma-t\bar Z_\gamma\right)
\cX_\gamma^{\rm sf}(t)\cX_{\gamma'}^{\rm sf}(t') + (\gamma \leftrightarrow \gamma')
\label{2instZ}
\ee
Defining $\psi_\gamma=\arg Z_\gamma$, $\psi_{\gamma\gamma'}= \psi_\gamma-\psi_{\gamma'}$ and changing the integration variables to 
$t= \I e^{\I\psi_\gamma} e^{\xp+\xm}$, $t'= \I e^{\I\psi_{\gamma'}} e^{\xp-\xm}$, this can be
rewritten as
\be
\begin{split}
\cI^{(2)}_{\gamma,\gamma'}=&\,\frac{\I R}{16\pi^3}
(-1)^{\langle \gamma,\gamma'\rangle} \,  \langle \gamma,\gamma'\rangle\, 
\bar\Omega(\gamma)\, \bar\Omega(\gamma') \,
\sigma_{\gamma+\gamma'} 
\int\limits_{-\infty}^\infty \de \xp \,\int\limits_{-\infty}^\infty \de \xm\, 
\coth\(\xm+\frac{\I}{2}\psi_{\gamma\gamma'}\)
\\
 \times&
\big[  |Z_\gamma|\cosh(\xp+\xm) +  |Z_{\gamma'}|\cosh(\xp-\xm) \big]\,
e^{-2\pi R|Z_\gamma|\cosh(\xp+\xm)-2\pi R|Z_{\gamma'}|\cosh(\xp-\xm)-2\pi\I\langle \gamma+\gamma', C\rangle}.
\label{2instZ2}
\end{split}
\ee
We shall be interested in the behavior of  $\cI^{(2)}_{\gamma,\gamma'}$ in the vicinity of the
wall of marginal stability where $\psi_{\gamma,\gamma'}\to 0$, in the limit $R\to\infty$. We assume
that $\gamma$ and $\gamma'$ are primitive vectors generating the positive cone of BPS states whose central charges
align at the wall, so that $\bar\Omega(\gamma,u)$ and $\bar\Omega(\gamma',u)$ are constant across 
the wall, and equal to $\Omega(\gamma,u)$ and $\Omega(\gamma',u)$. 
Away from the wall, the integrals over $x_+$ and $x_-$ are dominated by saddle points at  $x_+=x_-=0$, producing\footnote{The analysis in this section bears some similarities with the one in \cite{Chen:2010yr}.} 
\be
\cI^{(2)}_{\gamma,\gamma'}\approx \frac{ 1}{32\pi^3}
(-1)^{\langle \gamma,\gamma'\rangle} \,  \langle \gamma,\gamma'\rangle\, 
\bar\Omega(\gamma)\bar\Omega(\gamma')\,
\sigma_{\gamma+\gamma'}
\frac{ |Z_\gamma|+|Z_{\gamma'}|}{|Z_\gamma| \, |Z_{\gamma'}|}
\cot\(\hf\psi_{\gamma\gamma'}\) 
\,
e^{-2\pi R ( |Z_\gamma| + |Z_{\gamma'}| ) -2\pi\I \langle \gamma+\gamma' , C\rangle}.
\label{2instZ3}
\ee
However, the saddle point approximation breaks down near the wall where
$\psi_{\gamma\gamma'}\to 0$, as the saddle point
collides with the pole at $x_-=-\tfrac{\I}{2} \psi_{\gamma\gamma'}$. To deal with this, 
we first  perform the integral over $x_+$, which is dominated
by a saddle point at 
\be
x_+ \sim \frac{|Z_{\gamma'}|-|Z_{\gamma}|}{|Z_{\gamma}|+|Z_{\gamma'}|} x_- + \cO(x_-^2)\ .
\ee
In the limit $R\to\infty$, $\cI^{(2)}_{\gamma,\gamma'}$ is well approximated by 
\be
\begin{split}
\cI^{(2)}_{\gamma,\gamma'}\approx&\,\frac{\I}{16\pi^3}
(-1)^{\langle \gamma,\gamma'\rangle}  \,\langle \gamma,\gamma'\rangle\,
\bar\Omega(\gamma)\, \bar\Omega(\gamma') \,
\sigma_{\gamma+\gamma'} \sqrt{R (|Z_{\gamma}|+|Z_{\gamma'}|)}
\\
&\, \times \int\limits_{-\infty}^\infty \de \xm\,
\, 
\coth\(\xm+\frac{\I}{2}\,\psi_{\gamma\gamma'}\)
e^{-2\pi R ( |Z_\gamma|+|Z_{\gamma'}| ) -\frac{4\pi R |Z_\gamma| |Z_{\gamma'}|}
{ |Z_\gamma|+|Z_{\gamma'}| } x_-^2-2\pi\I\langle \gamma+\gamma', C\rangle}.
\end{split}
\ee
In the limit $\psi_{\gamma\gamma'}\to 0$, we can further approximate
$\coth(x)\sim 1/x$ and evaluate the integral using the formula \cite[4.18]{Alexandrov:2012au},
valid for $\alpha$ and $\beta$ real and non-zero,
\be
\int_{-\infty}^{\infty} \frac{\de z}{z-\I \alpha} e^{-\beta^2 z^2}
=\I \pi \, \sgn(\alpha)\, e^{\alpha^2\beta^2} {\rm Erfc}( |\alpha \beta|)
\ee
where ${\rm Erfc}$ is the complementary error function. 
Noting that 
\be
\label{diffZ}
  |Z_\gamma|+|Z_{\gamma'}| - |Z_{\gamma+\gamma'}| \sim \frac12 
  m_{\gamma\gamma'} \, 
   \psi_{\gamma\gamma'}^2\ ,
\ee
where $m_{\gamma\gamma'} =  \frac{|Z_\gamma|\, |Z_{\gamma'}| }{ |Z_\gamma|+|Z_{\gamma'}| }$
is the reduced mass of the two-particle system,
we find
\be
\label{Z2wall}
\begin{split}
\cI^{(2)}_{\gamma,\gamma'} \approx&\,\frac{1}{16\pi^2}
(-1)^{\langle \gamma,\gamma'\rangle}  
\bar\Omega(\gamma)\bar\Omega(\gamma') \langle \gamma,\gamma'\rangle
\sigma_{\gamma+\gamma'} 
\sqrt{R \(  |Z_\gamma|+|Z_{\gamma'}|  \)}  \\
\times &
\sgn( \psi_{\gamma\gamma'})\, 
{\rm Erfc}\left( |\psi_{\gamma\gamma'}| \sqrt{\pi R\, m_{\gamma\gamma'}} \right)\, e^{-2\pi R |Z_{\gamma+\gamma'}| -2\pi\I\langle \gamma+\gamma', C\rangle}
\end{split}
\ee
The two-particle contribution is discontinuous across the wall:  as $\psi_{\gamma\gamma'}$ goes from negative to positive, $\cI^{(2)}_{\gamma,\gamma'}$ jumps by 
\be
\Delta \cI^{(2)}_{\gamma,\gamma'} \approx \frac{1}{8\pi^2}
(-1)^{\langle \gamma,\gamma'\rangle}  
\bar\Omega(\gamma)\bar\Omega(\gamma') \langle \gamma,\gamma'\rangle
\sigma_{\gamma+\gamma'} 
\sqrt{R \(  |Z_\gamma|+|Z_{\gamma'}|  \)} \, e^{-2\pi R |Z_{\gamma+\gamma'}| -2\pi\I\langle \gamma+\gamma', C\rangle}\ .
\label{D2instZ}
\ee

On the other hand, the one-particle contribution $\cI^{(1)}_{\gamma,\gamma'}$ is also discontinuous
across the wall, due to the fact that the one-particle index $ \bar\Omega(\gamma+\gamma')$ 
jumps~\cite{Denef:2007vg}:\footnote{Here $\Theta(x)$ denotes the Heaviside step function, equal to 1 when $x>0$ and $0$ otherwise.}
\be
\label{bOmwall}
 \bar\Omega(\gamma+\gamma',u) =  \bar\Omega^+(\gamma+\gamma')- (-1)^{\langle \gamma,\gamma'\rangle}  \,
|\langle \gamma,\gamma'\rangle| \, 
\bar\Omega(\gamma)\, \bar\Omega(\gamma') 
\, \Theta(  \langle \gamma,\gamma'\rangle\, \psi_{\gamma\gamma'} )\ .
\ee
The first term in \eqref{bOmwall} corresponds to the one-particle index on the side  where
$ \langle \gamma,\gamma'\rangle\, \psi_{\gamma\gamma'}<0$, so that the two states of charge $\gamma$ and $\gamma'$ cannot form a BPS bound state, while the second term is the contribution 
of the BPS bound state which exists on the side where $\langle \gamma,\gamma'\rangle\, \psi_{\gamma\gamma'}>0$. 

Inserting \eqref{bOmwall} in
$ \cI^{(1)}_{\gamma+\gamma'}$, and taking the  limit $R\to \infty$, we find
\be
\label{Z1wall}
 \begin{split}
 \cI^{(1)}_{\gamma+\gamma'} \approx & 
 \left[ \bar\Omega^+(\gamma+\gamma') - (-1)^{\langle \gamma,\gamma'\rangle}  \,
|\langle \gamma,\gamma'\rangle| \, 
\bar\Omega(\gamma)\, \bar\Omega(\gamma') 
\Theta(  \langle \gamma,\gamma'\rangle\, \psi_{\gamma\gamma'} )
\right]\\
&\times
\sigma_{\gamma+\gamma'}  \frac{\sqrt{R |Z_{\gamma+\gamma'}|}}{8\pi^2} 
\,
e^{-2\pi R |Z_{\gamma +\gamma'}|-2\pi\I\langle \gamma+\gamma', C\rangle}\ ,
\end{split}
\ee
whose jump exactly compensates \eqref{D2instZ}. In fact, neglecting the difference between
$ |Z_{\gamma+\gamma'}|$ and $ |Z_{\gamma}|+|Z_{\gamma'}|$ under the square root (which cannot be told apart in our approximation), the sum of \eqref{Z2wall} and \eqref{Z1wall} can be written as 
\be
\label{Z12wall}
\begin{split}
\cI^{(1)}_{\gamma+\gamma'}+ \cI^{(2)}_{\gamma,\gamma'} \approx &
 \Big\{ \bar\Omega^+(\gamma+\gamma')  \\
 & - \frac12 (-1)^{\langle \gamma,\gamma'\rangle}  \,
|\langle \gamma,\gamma'\rangle| \, 
\bar\Omega(\gamma)\, \bar\Omega(\gamma') 
\left( 1 + {\rm Erf} \left(  \sgn(\langle \gamma,\gamma'\rangle)\, \psi_{\gamma\gamma'}
\sqrt{  \pi R \, m_{\gamma\gamma'} } \right) \right)
\Big\} \\
&\times
\sigma_{\gamma+\gamma'}  \frac{\sqrt{R |Z_{\gamma+\gamma'}|}}{8\pi^2} 
\,
e^{-2\pi R |Z_{\gamma +\gamma'}|-2\pi\I\langle \gamma+\gamma', C\rangle}\ ,
\end{split}
\ee
where we have used the identity ${\rm Erf}(x)=\sgn(x)\, (1-{\rm Erfc}(|x|)$. In plain words,
the addition of the two-particle contribution to $\cI^{(1)}_{\gamma+\gamma'}$ has converted the step
function $\Theta(x)$ in \eqref{bOmwall} into the smooth function $\frac12[1+{\rm Erf}(x)]$. 
This shows that the sum of the one and two-particle contributions  is not only continuous, but also differentiable across the wall (see Figure 1 for illustration), which acquires a finite width of order 
$1/\sqrt{R \, m_{\gamma,\gamma'}}$ as a function of the relative phase $\psi_{\gamma\gamma'}$
between the central charges $Z_\gamma$ and $Z_{\gamma'}$.
It would be interesting to generalize this computation to the case
of non-primitive wall-crossing, and to relax the non-relativistic limit $R\to\infty$.

\EPSFIGURE{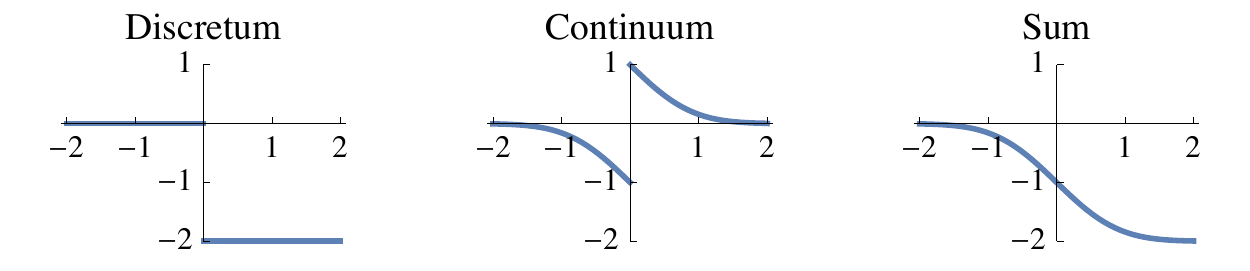,height=3.7cm}{Behavior of the one-particle contribution ($-2\Theta(x)$), two-particle 
contribution (${\rm sign}(x)\, {\rm Erfc}(|x|)$) and their sum ($-1-{\rm Erf}(x)$) to the index with total charge $\gamma+\gamma'$ across a wall where the phases of $Z_\gamma$ and $Z_{\gamma'}$ align ($x\to 0$). }

\section{Supersymmetric  electron-monopole quantum mechanics}

Our goal in the remainder of this note is to derive the two-particle contribution \eqref{Z2wall} from the supersymmetric quantum mechanics of a system of two non-relativistic particles with mutually non-local primitive charges $\gamma,\gamma'$. 
After factoring out the center of mass degrees of freedom, which can be treated as in \eqref{Z1vsTr}, 
and the internal degrees of freedom, counted by $\bar\Omega(\gamma) \bar\Omega(\gamma')$,
the system is described by 
$\cN=4$ quantum mechanics with Hamiltonian \cite{D'Hoker:1985kb,Denef:2002ru}\footnote{$\cN=4$
supersymmetry allows a position-dependent rescaling of the kinetic term \cite{Lee:2011ph}, but the spectral asymmetry is independent of this deformation, as long as it goes to one at spatial infinity.}
\be
\label{Hmag}
H=\frac{1}{2m}\,(\vec p-q \vec A)^2-\frac{q}{2m}\, \vec B\cdot \vec\sigma\otimes(1_2-\sigma_3)
+ \frac{1}{2m} \left( \vartheta - \frac{q}{r} \right)^2 .
\ee
where $q=\frac12\langle \gamma,\gamma'\rangle$ is half the Dirac-Schwinger-Zwanziger
product of the electromagnetic charges, $\vec B= \frac{\vec r}{r^3}$ is the magnetic field
of a unit charge magnetic monopole sitting at the origin, $\vec A$ is the corresponding gauge potential, $\vec\sigma$ are the Pauli matrices,
and $m= m_{\gamma\gamma'}$ is the reduced mass of the two-particle system.  Classically, the system has bound states for $q\vartheta>0$, no bound states for $q\vartheta<0$, and a continuum of scattering states 
with energy $E\geq  E_c = \frac{\vartheta^2}{2m}$. The  parameter $\vartheta$ is fixed by equating
$E_c$ with the binding energy,
\be
\frac{\vartheta^2}{2m} = |Z_\gamma| + |Z_{\gamma'}| - |Z_{\gamma+\gamma'}|,
\ee
so $\vartheta\sim m\psi_{\gamma\gamma'}$ near the wall, cf. \eqref{diffZ}. Quantum mechanically, 
 $H$ describes  two bosonic degrees of freedom with helicity $h=0$, and one fermionic doublet with helicity $h=\pm 1/2$ and gyromagnetic ratio $g=4$. This unusual value is fixed by the requirement
of supersymmetry, and can be understood as the combined effect of electromagnetic and scalar
interactions \cite{Horvathy:2005pj}. Indeed, the 
Hamiltonian \eqref{Hmag} commutes with the four supercharges 
(here $\vec \Pi = \vec p-q \vec A$) \cite{D'Hoker:1985kb,Denef:2002ru,Lee:2011ph}
\be
Q_4=\frac{1}{\sqrt{2m}} 
\begin{pmatrix} 0 & 
- \I \left( \vartheta - \frac{q}{r} \right)+\vec\sigma\cdot \vec \Pi
\\
\I \left( \vartheta - \frac{q}{r} \right) + \vec\sigma\cdot \vec \Pi & 
0 \end{pmatrix} \, 
\ee
\be
Q_a=\frac{1}{\sqrt{2m}} \begin{pmatrix} 0 &  -\left( \vartheta - \frac{q}{r} \right) \sigma_a - \I \Pi_a  +  \epsilon_{abc} \Pi_b \sigma_c \\
-\left( \vartheta - \frac{q}{r} \right)  \sigma_a + \I \Pi_a+ \epsilon_{abc} \Pi_b \sigma_c  & 0\end{pmatrix} .
\ee
which satisfy the algebra (here $m=1,2,3,4$)
\be
\{ Q_m, Q_n \} = 2 H\, \delta_{mn}\ .
\ee
The complete spectrum of this Hamiltonian was analyzed in \cite{Avery:2007xf}, but unfortunately
these authors stopped short of computing the density of states in the continuum. We shall revisit
this computation, adapting the classic treatment of the electron-monopole system without potential in \cite{Kazama:1976fm}.

The Hamiltonian \eqref{Hmag} commutes with the total angular momentum operator \cite{Kazama:1976fm}
\be
\vec J = \vec r \wedge (\vec p-q \vec A) - q\, \frac{\vec r}{r} + \frac14\, \vec \sigma \otimes (1_2-\sigma_3),
\qquad
[J_a,J_b]=\I \eps_{abc} J_c.
\ee
The Schr\"odinger equation $H \Psi = E \Psi$ can be solved by separating the angular and radial dependence.
For this we diagonalize $J_3$ and $\vec J^2$ denoting by $m$ and $j(j+1)$ their eigenvalues.
For the spin 0 part (corresponding to the first two entries of the eigenvector $\Psi$), we write
\be
\Psi =  f(r)
\, Y_{q,j,m}\ ,\quad j\geq |q|\, ,
\qquad
j-q\in \IZ\, ,
\ee
where $Y_{q,l,m}$ are the monopole harmonics (also known as spin-weighted spherical harmonics),
given in the patch around $\theta=0$ by \cite{Wu:1976ge} 
\be
Y_{q,l,m}=2^m \sqrt{\frac{(2l+1)\, (l-m)! (l+m)!}{4\pi\, (l-q)!\, (l+q)!}}
(1-\cos\theta)^{-\tfrac{q+m}{2}}(1+\cos\theta)^{\tfrac{q-m}{2}}\,
P_{l+m}^{-q-m,q-m}(\cos\theta)\, e^{\I(m+q)\phi}\, .
\ee
Here $P_n^{\alpha,\beta}(x)$ are the Legendre polynomials
\be
P_{n}^{\alpha,\beta}(x)= \frac{(-1)^n}{2^n n!} (1-x)^{-\alpha} (1+x)^{-\beta}
\frac{\de^n}{\de x^n} \left[ (1-x)^{\alpha+n} (1+x)^{\beta+n}\right].
\ee
Using
\bea
&&
(\vec p-q \vec A)^2=-\frac{1}{r}\pa_r^2\, r+\frac{1}{r^2}\left( \vec r \wedge (\vec p - q \vec A) \right)^2
\\
&&
\left( \vec r \wedge (\vec p - q \vec A) \right)^2 =\vec L^2 -q^2=
-\frac{1}{\sin\theta}\pa_\theta \, \sin\theta\, \pa_\theta -\frac{1}{\sin^2\theta}
\left(\pa_\phi+\I q (\cos\theta-1) \right)^2\ ,
\eea
one can show that 
the radial wave function satisfies \cite{Wu:1976ge}
\be
\label{Hschr0}
\left[ -\frac{1}{2m} \frac{1}{r} \pa_r^2\,  r + \frac{\nu^2-q^2-\tfrac14}{2mr^2} +
\frac{1}{2m}\left( \vartheta - \frac{q}{r} \right)^2- E \right]\, f(r) = 0\, ,
\ee
where $\nu=j+\half$. 

For the spin 1/2 part (corresponding to the last two entries of the eigenvector $\Psi$), the angular dependence is a linear combination
of modes with orbital momentum $j-\tfrac12$ and $j+\tfrac12$  \cite{Kazama:1976fm},
\be
\phi_{j,m}^{(1)}=\begin{pmatrix}
\sqrt{\frac{j+m}{2j}} Y_{q,j-\tfrac12, m-\tfrac12} \\
\sqrt{\frac{j-m}{2j}} Y_{q,j-\tfrac12, m+\tfrac12}
\end{pmatrix}\ ,\quad
\phi_{j,m}^{(2)}=\begin{pmatrix}
-\sqrt{\frac{j-m+1}{2j+2}} Y_{q,j+\tfrac12, m-\tfrac12} \\
\sqrt{\frac{j+m+1}{2j+2}} Y_{q,j+\tfrac12, m+\tfrac12}
\end{pmatrix}\ ,\quad j-q\in \IZ+\frac12
\ee
The first set of modes occurs for $j\geq |q|+\hf$ while the second
occurs for $j\geq |q|-\hf$. In order to diagonalize the action of $\vec\sigma\cdot \vec r$ and
$\vec\sigma\cdot(\vec p-q \vec A)$, which commute with $\vec J$, it is convenient to introduce
the linear combinations (for $j\geq |q|+\hf$)  \cite{Kazama:1976fm}
\be
\xi^{(+)}_{j,m} = c_+ \, \phi^{(1)}_{j,m} -c_- \, \phi^{(2)}_{j,m} ,
\qquad
\xi^{(-)}_{j,m} = c_-  \, \phi^{(1)}_{j,m} + c_+ \, \phi^{(2)}_{j,m}
\ee
where the coefficients
\be
c_\pm=\frac{q \left( \sqrt{2j+1+2q} \pm \sqrt{2j+1-2q}\right)}
{|q|\, \sqrt{2(4j+2)}}
\ee
satisfy $c_+^2+c_-^2=1$. Using $\vec r\cdot \vec\sigma = 2r\sqrt{\frac{\pi}{3}}
\begin{pmatrix} Y_{0,1,0} & \sqrt2 Y_{0,1,-1} \\ -\sqrt2 Y_{0,1,1} & -Y_{0,1,0} \end{pmatrix}$,
and the multiplication rule
\be
\begin{split}
Y_{q_1,j_1,m_1}\, Y_{q_2,j_2,m_2}=&
\sum_{j_3={\rm max}(|j_1-j_2|,|m_1+m_2|)}^{j_1+j_2}
\sqrt{\frac{(2j_1+1)(2j_2+1)}{4\pi(2j_3+1)}}\,
\langle j_1,-q_1,j_2,-q_2 | j_3,-q_1-q_2\rangle\, \\
& \times \langle j_1,m_1,j_2,m_2 | j_3,m_1+m_2\rangle\, Y_{q_1+q_2,j_3,m_1+m_2}
\end{split}
\ee
for monopole harmonics, 
one can show that these combinations satisfy, for any $f(r)$,
\be
\label{kazama1}
\begin{split}
(\vec\sigma\cdot\vec r)\, f(r) \xi^{(\pm)}_{j,m} =& - r \, f(r) \xi^{(\mp)}_{j,m},
\\
\vec\sigma\cdot (\vec p-q \vec A)\, f(r) \xi^{(\pm)}_{j,m} = &
\I \left(\pa_r+ r^{-1} (1\mp \mu) \right)\, f(r)\, \xi^{(\mp)}_{j,m},
\end{split}
\ee
where we defined
\be
\mu = \sqrt{(j+\tfrac12)^2 - q^2}  .
\ee
Using the fact that the Hamiltonian in the spin 1/2 sector can be written as
\be
\label{Hmaghalf1}
H_{1/2}= \frac{1}{2m}\( \vec \sigma\cdot (\vec p-q \vec A)\)^2-\frac{q}{2m}\, \vec B\cdot \vec\sigma
+ \frac{1}{2m} \left( \vartheta - \frac{q}{r} \right)^2
\ee
and the identity
\be
\left(\pa_r + \frac{1\pm \mu}{r}\right) \left(\pa_r + \frac{1\mp \mu}{r}\right) =
\frac{1}{r} \pa_r^2 r - \frac{\mu(\mu\mp 1)}{r^2},
\ee
we find that its action on $f(r)\xi^{(\pm)}_{j,m}$ is given by
\be
\begin{split}
H_{1/2}\cdot f(r)\, \begin{pmatrix}  \xi^{(+)}_{j,m} \\  \xi^{(-)}_{j,m}\end{pmatrix} = &
\left[ -\frac{1}{2m} \frac{1}{r} \pa_r^2\,  r + \frac{\mu^2}{2mr^2} +
\frac{1}{2m}\left( \vartheta - \frac{q}{r} \right)^2 \right]\,
\cdot f(r)\, \begin{pmatrix}  \xi^{(+)}_{j,m} \\  \xi^{(-)}_{j,m}\end{pmatrix}
\\
& +  \frac{1}{2mr^2} \begin{pmatrix} -\mu & q \\ q & \mu \end{pmatrix} \cdot
\,f(r)\, \begin{pmatrix} \xi^{(+)}_{j,m} \\  \xi^{(-)}_{j,m}\end{pmatrix} .
\end{split}
\ee
The  $2\times 2$ matrix on the second line has eigenvalues $\pm\sqrt{\mu^2+q^2}=\pm(j+\tfrac12)$, and eigenvectors
\be
\label{xitpm}
\tilde\xi^{(\pm)}_{j,m} = (\mu\mp\sqrt{\mu^2+q^2}) \, \xi^{(+)}_{j,m} - q  \, \xi^{(-)}_{j,m}
\ee
Noting that the coefficient of the centrifugal $1/r^2$ term in the potential is proportional to
\be
\mu^2 \pm \sqrt{\mu^2+q^2} = \left( j + \frac12 \pm \frac12\right)^2 - q^2 - \frac14\, ,
\ee
we find that the radial wavefunctions $f_\pm(r)$  associated to the eigenmodes
$\tilde\xi^{(\pm)}_{j,m}$
satisfy the same equation as \eqref{Hschr0} with $\nu=j+1$ (for the $+$ sign,
which we refer to as the helicity $h=\half$ mode) or $\nu=j$  (for the $-$ sign, which we refer to as the helicity $-\half$ mode).

Finally, for $j= |q|-1/2$, the space of eigenmodes of $J^2, J_3$ is one-dimensional,
spanned by
\be
\eta_m \equiv \phi^{(2)}_{j,m} \propto (1+\cos\theta)^{\tfrac{j-m}{2}} (1-\cos\theta)^{\tfrac{j+m-1}{2}}\,
\e^{\I (m+j)\phi}\, \begin{pmatrix} \sin\theta \\ e^{\I\phi} (1- \cos\theta) \end{pmatrix}   .
\ee
One has, in place of \eqref{kazama1},
\be
\label{kazama2}
(\vec\sigma\cdot\vec r)\, \eta_m = r \frac{q}{|q|} \eta_m\, ,
\qquad
\vec\sigma\cdot (\vec p-q \vec A)\, f(r) \,\eta_m = -\I \frac{q}{|q|}
(\pa_r + r^{-1})\, f(r)\, \eta_m\, ,
\ee
leading to the same equation \eqref{Hschr0} with $\nu=j$.

In summary, the radial equation is given by \eqref{Hschr0} with 
\be
\nu=j+h+\tfrac12\ ,\quad j=|q|+h+\ell
\ee
with $h=0$ for the two bosonic degrees of freedom and $h=\pm \half$
for the spin $1/2$ degree of freedom,
and $\ell\in \IN$ in all cases. Solutions to \eqref{Hschr0}
with energy $E\equiv \frac{k^2}{2m}>\frac{\vartheta^2}{2m}$ are linear combinations\footnote{It  helps to note that \eqref{Hschr0} is isomorphic to  the Schr\"odinger equation of the hydrogen atom, 
whose radial wave-functions are linear combinations of 
$M_{\I\eta, \ell+\tfrac12}(2\I k r)$ and $W_{\I\eta, \ell+\tfrac12}(2\I k r)$
where $\eta=q_1 q_2 m/k$, $E=k^2/2m$. (see e.g. \cite[Chap. 14.6]{Newton:1982qc}).}
\be
r f(r) = \beta \, M_{-\frac{\I q\vartheta}{\sqrt{k^2-\vartheta^2}},
\nu}\left(2\I r\sqrt{k^2-\vartheta^2}\right)
+ \gamma \, W_{-\frac{\I q\vartheta}{\sqrt{k^2-\vartheta^2}},\nu}\left(2\I r\sqrt{k^2-\vartheta^2}\right) ,
\ee
where $M_{\lambda,\nu}(z)$ and $W_{\lambda,\nu}(z)$ are Whittaker functions, which are solutions of the second order differential equation
\be
\label{whittakereq}
\cD_{\lambda,\nu}\cdot w(z) \equiv
\left[ \pa_z^2 -\frac14 + \frac{\lambda}{z}+\frac{\tfrac14-\nu^2}{z^2}\right]\, w(z)=0
\ee
satisfying
\be
\label{asymMW}
M_{\lambda,\nu}(z)\mathop{\sim}_{z\to 0} z^{\nu+\tfrac12} \, ,
\qquad
W_{\lambda,\nu}(z)\mathop{\sim}_{|z|\to\infty} z^{\lambda} e^{-z/2}\, .
\ee
The solution proportional to $M$ is regular at $r=0$, while the solution proportional to $W$ describes an outgoing spherical wave.
Since
\begin{equation}
\label{WtoM}
M_{\lambda,\nu}(z) = \frac{\varGamma(2\nu+1)}{\varGamma(\nu-\lambda+\tfrac12)}\,
e^{\I\pi\lambda}\, W_{-\lambda,\nu}(e^{\I\pi}z)+
\frac{\varGamma(2\nu+1)}{\varGamma(\nu+\lambda+\tfrac12)}
e^{\I\pi(\lambda-\nu-\frac12)}\, W_{\lambda,\nu}(z) \, ,
\end{equation}
we find that the S-matrix in angular momentum channel $\ell$ and helicity $h$ is
\be
S_{h,\ell}(k)=
\frac{\Gamma(\nu+\lambda+\frac12)}{\Gamma(\nu-\lambda+\frac12)} = 
\frac
{\Gamma\left( |q| + \ell+2h+1
- \I \frac{q\vartheta}{\sqrt{k^2-\vartheta^2}}\right)}
{\Gamma\left( |q| + \ell+2h+1
+ \I \frac{q\vartheta}{\sqrt{k^2-\vartheta^2}}\right)}\ .
\ee
In particular, bound states  correspond to poles of the S-matrix, and occur only when $q\vartheta>0$, with energy
\be
E_{h,\ell,n}
=  \frac{\vartheta^2}{2m} \left( 1- \frac{q^2}{(|q|+\ell+2h+n+1)^2} \right)\ ,
\ee
with $n\geq 0$. The energy 
depends only on the sum $N=\ell+n+2h$, so the spectrum has additional degeneracies
beyond those predicted by rotational symmetry 
and supersymmetry \cite{D'Hoker:1985kb,Avery:2007xf}. The supersymmetric
ground state occurs in the  $h=-\half$ sector with $n=\ell=0$ and
has degeneracy $2|q| = |\langle\gamma_1,\gamma_2\rangle|$.
Its wave function is ${\scriptsize \begin{pmatrix} 0 \\   r^{q-1}\, e^{-\vartheta r}  \eta_m \end{pmatrix}}$, in agreement with \cite[4.16]{Denef:2002ru}.

The density of states (minus the density of states for a free particle in $\IR^3$) is the derivative of
the scattering phase, $\rho(k)\, \de k =  \frac{1}{2\pi\I} \de \log S(k)$.
The canonical partition function for states of helicity $h$ is therefore
\be
\begin{split}
 {\rm Tr}_{h} e^{-2\pi R H}=&\,
\Theta(q\vartheta)\, \sum_{\ell=0}^{\infty} \, \sum_{n=0}^{\infty} (2\ell+2|q|+2h+1)\, e^{-2\pi R E_{h,\ell,n}}
\\
+& \sum_{\ell=0}^{\infty}  (2\ell+2|q|+2h+1)\,  \int_{k=|\vartheta|}^{\infty}\, \frac{\de k\, \pa_k}{2\pi\I}\, \log
\frac
{\Gamma\left(|q|+\ell+2h+1
- \I \frac{q\vartheta}{\sqrt{k^2-\vartheta^2}}\right)}
{\Gamma\left(|q|+\ell+2h+1
+ \I \frac{q\vartheta}{\sqrt{k^2-\vartheta^2}}\right)}\,
e^{-\frac{\pi R k^2}{m}},
\end{split}
\ee
where  the first term, corresponding to discrete bound states, contributes
only when $q\vartheta>0$. Summing over all types $h$ weighted by fermionic parity $(-1)^{2h}$,
only the BPS bound state with $E=0$ contributes from the first term, while the contribution of the
continuum of scattering states simplifies to
\be
\label{Trcancel}
\begin{split}
\sum_{\ell=0}^{\infty}
\int\limits_{k=|\vartheta|}^{\infty}\,  \frac{\de k\, \pa_k}{2\pi\I}\,
\left[
  (2\ell+2|q|)
  \log \frac{z_\ell}{\bar z_\ell}-   (2\ell+2|q|+2) \log \frac{z_{\ell+1}}{\bar z_{\ell +1}} \right]
  e^{-\frac{\pi R k^2}{m}},
\end{split}
\ee
where
\be
z_\ell = |q|+\ell- \I \frac{q\vartheta}{\sqrt{k^2-\vartheta^2}}\, .
\ee
Cancelling the terms in the sum,  only the contribution $\ell=0$ remains,
leading to
 \be
\begin{split}
\label{indexcont2}
2|q|\, \int\limits_{k=|\vartheta|}^{\infty}\,  \frac{\de k\, \pa_k}{2\pi\I} \log\left[
 \frac{ |q|- \I \frac{q\vartheta}{\sqrt{k^2-\vartheta^2}}}{ |q|+ \I \frac{q\vartheta}{\sqrt{k^2-\vartheta^2}}} \right] e^{-\frac{\pi R k^2}{m}}
= &\frac{2 q \vartheta}{\pi} \int\limits_{k=|\vartheta|}^{\infty}\, \frac{\de k}{k\sqrt{k^2-\vartheta^2}}  e^{-\frac{\pi R k^2}{m}} \ .
\end{split}
\ee
This is in fact the standard spectral asymmetry predicted by Callias' theorem \cite{Callias:1977kg}\footnote{While the Callias theorem is valid a smooth monopole background, the extension to 
singular monopoles was worked out in \cite{Moore:2014jfa}, 
and leads to the same spectral asymmetry. I thank
A. Royston and D. van den Bleeken for discussions on this matter.}.  Indeed, 
the result \eqref{indexcont2} does not depend on the details of the S-matrix, but only on the
ratio $S_{h,\ell+1}(k)/S_{h,\ell}(k)$, which as we discuss in an Appendix, is fixed by supersymmetry
in the asymptotic region.

Using
\be
\int\limits_{k=|\vartheta|}^{\infty}\,  \frac{\de k}{k \sqrt{k^2-\vartheta^2} }e^{-\frac{\pi R k^2}{m}} =
\frac{\pi}{2|\vartheta|} {\rm Erfc}\left( |\vartheta|\, \sqrt{\frac{\pi R}{m}} \right), 
\ee
and adding in the bound state contribution, we get, for arbitrary signs of $q$ and $\vartheta$,
\be
\begin{split}
{\rm Tr} (-1)^{F}\, e^{-\pi t H}=&\, -|2q|\, \Theta(q\vartheta) +  {\rm sign}(q\vartheta)\, |q|\, {\rm Erfc}\left( |\vartheta|\, \sqrt{\frac{\pi R}{m}} \right)\\
=& -q \left[ {\rm sign}(q)+  {\rm Erf}\left( \vartheta\, \sqrt{\frac{\pi R}{m}} \right) \right]
\end{split}
\ee
This is indeed a smooth function of $\vartheta$, which interpolates from $0$ at $\vartheta=-\infty$
to $-2q$ at $\vartheta=+\infty$ when $q>0$, or from $-2|q|$  at $\vartheta=-\infty$ to 
$0$ at $\vartheta=+\infty$ when $q<0$ (see Figure 1, which displays the case $q=1$). 
Including the degeneracy $\bar\Omega(\gamma) \bar\Omega(\gamma')$
of the internal degrees of freedom, and the contribution of the
center of motion degrees of freedom, given in the last line of \eqref{Z12wall},
we find perfect agreement with the  two-particle contribution to the index 
predicted by the formula \eqref{defZc}.

\section{Discussion}

In this note, I have shown that the general formula for the index \eqref{defZc} 
in $\cN=2$, $D=4$ gauge theories correctly reproduces the contribution of the continuum of two-particle states, in the vicinity of a wall of marginal stability where the constituents can be treated as non-relativistic BPS particles. In particular, I demonstrated that the contributions of the BPS bound states and of the two-particle continuum add up to a smooth function of the moduli, even though each of them is separately discontinuous across the wall. This analysis provides a physical justification for the replacement $\sgn(x)\to {\rm Erf}(x)$, which has been 
postulated in studies of black hole partition functions  in order to enforce S-duality or modular invariance \cite{Manschot:2009ia,Dabholkar:2012nd,Cardoso:2013ysa,Cardoso:2014hma}, a trick borrowed from the mathematics literature on indefinite theta series \cite{Zwegers-thesis}. It would be very interesting to calculate the contribution of the continuum of multi-particle states
away from the wall, a challenge that  will require to understand the dynamics of a collection of relativistic mutually non-local particles beyond the BPS regime.

It is worthwhile noting that similar smooth interpolations across walls of marginal stability have been encountered recently in localization computations of the index in gauged supersymmetric quantum mechanics  in certain scaling limits \cite{Hwang:2014uwa,Hori:2014tda}. More generally, error function profiles are ubiquitous in the context of Stokes phenomenon \cite{Berry1988}, which is formally similar with wall-crossing \cite{Bridgeland2012}. It would be interesting to explore these connections.

\medskip

\noindent {\bf Acknowledgments}: It is a pleasure to thank A. Neitzke, J. Manschot, G. Moore, A. Royston, A. Sen, J. Troost, D. van den Bleeken and P. Yi for  useful discussions. Special thanks are due to S. Alexandrov for  collaboration at an initial stage of this work and continued collaboration on closely related topics.

\appendix

\section{Robustness of the spectral asymmetry}

In order to elucidate the origin of the cancellations in \eqref{Trcancel}, we need to understand how 
supersymmetry relates the density
of states in the bosonic and fermionic sectors. For this purpose, notice that the operators
\be
\begin{split}
\label{Qradial}
Q_r=&\pa_z - \left[ \frac{\nu+\half}{z} - \frac{\lambda}{2\nu+1}\right] =
\frac{1}{2\I\sqrt{k^2-\vartheta^2}}\left[ \pa_r - \frac{\nu+\tfrac12}{r} + \frac{q\vartheta}{\nu+\tfrac12}
\right] \\
Q'_r=&\pa_z + \left[ \frac{\nu-\half}{z} - \frac{\lambda}{2\nu-1}\right] =
\frac{1}{2\I\sqrt{k^2-\vartheta^2}}\left[ \pa_r + \frac{\nu-\tfrac12}{r} - \frac{q\vartheta}{\nu-\tfrac12}
\right]
\end{split}
\ee
maps solutions of the Whittaker equation \eqref{whittakereq} with parameters $(\lambda,\nu)$ to solutions
of the same equation with parameters $(\lambda,\nu+1)$ and  $(\lambda,\nu-1)$,
\be
\label{QrcD}
Q_r\cdot \cD_{\lambda,\nu} =  \cD_{\lambda,\nu+1}\cdot Q_r\ ,\quad
Q'_r\cdot \cD_{\lambda,\nu} =  \cD_{\lambda,\nu-1}\cdot Q'_r\ .
\ee
 In fact, $Q_r$ and $Q'_r$ can be interpreted as the supercharge for the radial problem.  To see this, consider
acting with $\sqrt{2m} \, Q_{1/2}=\vec \sigma \cdot \vec\Pi+\I (\frac{q}{r}-\vartheta)$ 
on fermionic eigenfunctions
$f_\pm \,  \txi^{(\pm)}_{j,m}$: this should produce linear combinations of bosonic eigenfunctions with
the same energy and spin $j\pm \tfrac12$, namely $f_1\, \phi^{(1)}_{j,m}$ and $f_2 \, \phi^{(2)}_{j,m}$.
Here, $f_+, f_-, f_1,f_2$ are solutions of the radial equation \eqref{Hschr0} with $\nu=j+1, j, j, j+1$, respectively.
Indeed, we find
\be
\begin{split}
\I \sqrt{2m}\, Q_{4}\cdot f_+\, \txi^{(+)}_{j,m} = - (2j+1)\, c_-\, \left(
\pa_r + \frac{j+\tfrac32}{r} - \frac{q\vartheta}{j+\tfrac12 } \right) f_+ \, \phi^{(1)}_{j,m}
+ 2 \mu \vartheta c_-\,  f_+ \, \phi^{(2)}_{j,m} \\
\I \sqrt{2m}\, Q_{4}\cdot f_-\, \txi^{(-)}_{j,m} = - (2j+1)\, c_+\, \left(
\pa_r - \frac{j-\tfrac12}{r} + \frac{q\vartheta}{j+\frac12} \right) f_- \, \phi^{(2)}_{j,m}
- 2\mu \vartheta c_+ \,  f_- \, \phi^{(1)}_{j,m}
\end{split}
\ee
The differential operator in brackets coincides with $r^{-1} \cdot Q'_r\cdot r$ and $r^{-1} \cdot Q_r\cdot r$,
up to overall normalization.  Acting on the Whittaker wave-functions,
we have, as a consequence of \eqref{QrcD} and \eqref{asymMW}\footnote{The last equation in \eqref{QrWM} was noted in  \cite[VI.24]{Avery:2007xf}.},
\be
\label{QrWM}
\begin{split}
Q_r\cdot W_{\lambda,\nu}(z) =& \frac{\lambda-\nu-\tfrac12}{2\nu+1} W_{\lambda,\nu+1}(z)\\
Q_r\cdot W_{-\lambda,\nu}(-z) =& \frac{\lambda+\nu+\tfrac12}{2\nu+1} W_{-\lambda,\nu+1}(-z)\\
Q_r\cdot M_{\lambda,\nu}(z) =& \frac{(\nu+\tfrac12)^2-\lambda^2}{2(2\nu+1)^2(\nu+1)} M_{\lambda,\nu+1}(z)
\end{split}
\ee
It follows from these relations that the reflection coefficients $A(\lambda,\nu)$, $B(\lambda,\nu)$ defined by 
\be
M_{\lambda,\nu}(z) = A(\lambda,\nu)\,
 W_{-\lambda,\nu}(-z)+B(\lambda,\nu)
e^{\I\pi(-\nu-\frac12)}\, W_{\lambda,\nu}(z) \
\ee
satisfy
\be
\frac{A(\lambda,\nu+1)}{A(\lambda,\nu)}=\frac{2(\nu+1)(2\nu+1)}{\nu-\lambda+\tfrac12}\ ,\quad
\frac{B(\lambda,\nu+1)}{B(\lambda,\nu)}=\frac{2(\nu+1)(2\nu+1)}{\nu+\lambda+\tfrac12}\ .
\ee
Denoting  the reflection coefficient $S_\nu(\lambda)=A(\lambda,\nu)/B(\lambda,\nu)$, one has
\be
\label{Crecur}
\frac{S_{\nu+1}(\lambda)}{S_\nu(\lambda)} = \frac{\nu+\lambda+\tfrac12}{\nu-\lambda+\tfrac12}
\ee
This relation only depends on the  first two equations
in \eqref{QrWM}, which in turn follow directly from the action of $Q_r\sim\pa_r+\frac{\lambda}{2\nu+1}$ on the leading  asymptotic
behavior $W_{\lambda,\nu}\sim z^{\lambda} e^{-z/2}$ of the incoming/outgoing plane waves.
More generally, the ratio \eqref{Crecur} depends only on the supercharge at radial infinity and should
be unaffected by short-distance corrections to the potential or to the conformal factor in the metric.


\providecommand{\href}[2]{#2}\begingroup\raggedright\endgroup


\end{document}